\documentclass[%
 reprint,
 amsmath,amssymb,
 aps,
]{revtex4-2}
\usepackage{tikz}
\usepackage{lipsum}
\usepackage{graphicx}
\usepackage{dcolumn}
\usepackage{bm}

\begin{document}

\preprint{APS/123-QED}

\title{Particle Deflection around Microscopic Loop Quantum Black Hole}

\author{Haida Li}
\email{eqwaplay@scut.edu.cn}
\affiliation{School of Physics and Optoelectronics, South China University of Technology, Guangzhou 510641, China}

\author{Xiangdong Zhang} 
\email{Corresponding author: scxdzhang@scut.edu.cn}
\affiliation{School of Physics and Optoelectronics, South China University of Technology, Guangzhou 510641, China}

\begin{abstract}
The detection of quantum gravity effects is highly limited in both macroscopic and microscopic scenarios: The small quantum parameter makes most large-scale observations practically indistinguishable from general relativity. While at the Planck scale, where the effect of quantum gravity is undoubtedly significant, the energy requirement is remarkably high for any test particle as a probe. In this work, by focusing on the inner-most stable circular orbit (ISCO) of both massless and massive particles around a microscopic loop quantum black hole, we show that given the current knowledge of ultra-high-energy particles, quantum corrections of relative scale $>10^{-10}$ can appear when the radius of the black hole horizon remains larger than the wavelength of high-energy Gamma rays, where the classical photon trajectory may still hold to a certain degree. In addition, more significant corrections can be observed when considering low velocity massive test particles. Potentially, our results could suggest a new area for testing the effects of quantum gravity.
\end{abstract}

\maketitle


\section{\label{sec:111}Introduction}

Quantum gravity seeks to reconcile quantum mechanics and general relativity in extreme regimes such as black hole singularities and the early universe. While a complete quantum theory of gravity is very tantalizing, the absence of empirical tests due to Planck-scale effects makes the experimental search for quantum gravity extremely difficult.

Loop Quantum Gravity (LQG) is a background-independent candidate theory of quantum gravity \citep{Thiemann:2007pyv,Rovelli:2014ssa,Ashtekar:2017yom}. Its symmetry-reduced models, among which the most famous are Loop Quantum Cosmology (LQC) \citep{Ashtekar:2011ni} and Loop Quantum Black Hole (LQBH) \citep{Modesto:2005zm,Ashtekar:2018lag,Gambini:2020nsf,Kelly:2020uwj,Han:2020uhb,Lewandowski:2022zce,Zhang:2023yps,Giesel:2023tsj,Lin:2024flv,Zhang:2023yps} models, have been studied extensively, providing effective dynamics of the theory that can be directly compared with classical general relativity (GR). The most important result is the resolution of big bang singularity by big bounce \citep{Ashtekar:2006rx}, and the resolution of BH singularity.

So far, there have been some proposals to the observable quantum effects of LQG. For example, using Cosmological Microwave Backgound (CMB) to test the effects of LQC \citep{Wilson-Ewing:2016yan,Li:2021mop}, and white hole (WH) signals in our universe originating from the pre-bounce companion black hole (BH) during BH-WH transition \citep{Zhang:2023okw}.

The strong gravitational lensing near the BH \citep{Virbhadra:1999nm,Bozza:2002zj,Virbhadra:2008ws,Tsukamoto:2016qro,Igata:2025taz} may also serve as a potential testing ground. Particularly, gravitational lensing effects from LQG have been studied by many works \citep{Sahu:2015dea,Fu:2021fxn,Kumar:2023jgh,Junior:2023xgl}, most of these works treat the quantum parameter as a running parameter valued at order $1$. In a recent work \citep{Li:2024afr}, the quantum corrections of LGBH models with rigorously imposed quantum parameters on large-scale gravitational lensing effects are investigated by keeping strictly the Planck scale quantum parameters as originally proposed in the theory. The result shows straightforwardly that the quantum correction of LQG at a large scale is marginal, with the largest correction coming from the Ashtekar-Olmedo-Singh (AOS) model \citep{Ashtekar:2018cay}, at $\sim10^{-30}$ order. Meanwhile, as has been demonstrated clearly in some recent works \citep{del-Corral:2022kbk,Motaharfar:2025typ}, the quantum effect of LQG is expected to be larger as the black hole mass approaches the Planck mass ($M_{\mathrm{Pl}}=2.176\times 10^{-8}\ \mathrm{kg}$). This hints us to look at the microscopic black holes.

In this work, we focus on one of the most essential properties of a black hole: the innermost stable circular orbit (ISCO). By fixing the center black hole, the ISCO depends only on the properties of specific test particles (their mass, initial velocity, etc.). The ISCO, as observed angularly by a distant observer, is also a gravitational lens observable, which in many cases can have a level of quantum corrections similar to other lens observables, as shown in \citep{Li:2024afr}. We believe the results we obtained for the quantum corrections of ISCO can also be applied to other observables.

Additionally, some fundamental restrictions need to be considered for detecting the quantum correction of LQG. First, it is crucial that the characteristic scale of the black hole (where the effects of quantum gravity can be significant enough), namely the size of its horizon, is larger than the wavelength of the test particle. Otherwise, the particle's trajectory around the black hole will be impacted by its diffraction and corrections from quantum field theory in curved spacetimes. Second, since the black holes we study in this work are microscopic (typically with $r_H=10^{-20}\mathrm{m}$), we also need to consider the rate at which such black holes evaporate. Within the observation period, if the black hole shrinks by a decent amount, then the impact on its gravity may offset any observational effect of quantum gravity. Finally, observer motion during the observation period can also create errors that can potentially exceeds the level of quantum correction, we will provide an simple estimation of this effect as well.

Also, as quantum gravity remains an open question, we would like to keep our results as general as possible so that our results not only reflect on the quantum correction of one particular LQG model at a certain scale but also can be extended to other theories of quantum gravity. Since the framework we use to compute the ISCO and the observed angle by an observer is naturally generalized in the case of spherically symmetric spacetime, and also because the level of quantum corrections in many cases is straightforwardly proportional to the quantum correction of the metric tensor(see e.g. \citep{Li:2024afr}). We believe the results we obtained in this work can be extended to other black hole models of quantum gravity with similar levels of quantum correction to the metric.

The structure of this work is as follows. In Section \ref{sec2}, we briefly introduce the theoretical background of the calculations we do in this work. This includes a description of the AOS and qOS model considered, the computation of the ISCO of both massless and massive particles, and how to compute the observed angle of the ISCO for a free-falling observer traversing the deflection event plane (the equatorial plane). In Section \ref{sec3}, several key results will be presented, including both the absolute and relative difference in the observed angle of ISCO for massless particle (photon) and massive particle, the black hole evaporation time and the observer motion as two additional constraints, and also the trend in which the quantum correction increases as the black hole mass decreases all the way to the Planck scale. Finally, we will provide some discussion on the results we obtained in this work in Section \ref{sec4}.

\section{Theoretical Background}\label{sec2}

In this paper, we focus on the static spherically symmetric spacetimes obtained from the effective models of LQG, which in general can be described by line element:
\begin{equation}\label{generalG}
\mathrm{d} s^2=-A(r) \mathrm{d} t^2+B(r) \mathrm{d} r^2+C(r) \mathrm{d} \Omega^2.
\end{equation}
Specifically, we consider the region outside of the BH horizon of the Ashtekar-Olmedo-Singh (AOS) model \citep{Ashtekar:2018cay,Ashtekar:2020ckv,Zhang2020a}, which has the overall largest quantum corrections that have been reported in \citep{Li:2024afr}:

\begin{widetext}
\begin{equation}\label{AOS}
\begin{aligned}
& A_{A O S}(r)=\left(\frac{r}{r_H}\right)^{2 \epsilon} \frac{\left(1-\left(\frac{r_H}{r}\right)^{1+\epsilon}\right)\left(2+\epsilon+\epsilon\left(\frac{r_H}{r}\right)^{1+\epsilon}\right)^2\left((2+\epsilon)^2-\epsilon^2\left(\frac{r_H}{r}\right)^{1+\epsilon}\right)}{16\left(1+\frac{\delta_c^2 L_0^2 \gamma^2 r_H^2}{16 r^4}\right)(1+\epsilon)^4}, \\
& B_{A O S}(r)=\left(1+\frac{\delta_c^2 L_0^2 \gamma^2 r_H^2}{16 r^4}\right) \frac{\left(\epsilon+\left(\frac{r}{r_H}\right)^{1+\epsilon}(2+\epsilon)\right)^2}{\left(\left(\frac{r}{r_H}\right)^{1+\epsilon}-1\right)\left(\left(\frac{r}{r_H}\right)^{1+\epsilon}(2+\epsilon)^2-\epsilon^2\right)},\qquad C_{A O S}(r)=r^2\left(1+\frac{\gamma^2 L_0^2 \delta_c^2 r_H^2}{16 r^4}\right),
\end{aligned}
\end{equation}
\end{widetext}
for $r\in [r_H,\infty)$, $r_H=2GM$ is the Schwarzschild radius, and:
\begin{equation}
L_o \delta_c=\frac{1}{2}\left(\frac{\gamma \Delta^2}{4 \pi^2 M}\right)^{1 / 3}, \quad \epsilon+1=\sqrt{1+\gamma^2\left(\frac{\sqrt{\Delta}}{\sqrt{2 \pi} \gamma^2 M}\right)^{2 / 3}},
\end{equation}
are quantum parameters depending on both the Imirzi parameter $\gamma$ and the area gap $\Delta=21.17l_p^2$ (when setting $\gamma=1$), namely the minimum nonzero eigenvalue of the area operator in LQG. $l_p$ is Planck length. 

One easy way to evaluate the change of quantum correction with respect to the scale (mass) of the black hole is by performing the rescaling $r_H=\frac{2GM}{c^2}\rightarrow r'=r\cdot\frac{c^2}{G M}=2$. Then $l_p$ rescales as:
\begin{equation}\label{lpm}
    l_p\rightarrow l_p'=l_p\cdot \frac{c^2}{G M},
\end{equation}
where $c$ is the speed of light, $G$ is the gravitational constant, and $M$ is the actual mass of the black hole. It is straightforward under this rescaling to see that as the actual mass of the black hole decreases, the rescaled quantum parameter $l_p'$ becomes larger, generating larger quantum corrections.

In this work, we focus on test particles' innermost stable circular orbit (ISCO). The calculations of ISCO for both massless and massive particles in static spherically symmetric spacetimes are quite straightforward. (also for the innermost stable orbit of stationary axisymmetric spacetimes in the equatorial plane, see \citep{Duan:2023gvm}) First, we begin with the following geodesic equations of particle trajectory:
\begin{equation}
\begin{aligned}
\dot{\phi} & =\frac{L }{C},\ \dot{t}=\frac{E}{A},\ \dot{r}^2 =\frac{  C\left(E^2-\kappa A\right)- L^2 A}{ A B C},
\end{aligned}
\end{equation}
where $\kappa=1$ corresponds to a massive particle and $\kappa=0$ corresponds to a massless particle, $E$ and $L$ are the energy and angular momentum of the test particle per unit mass, the dots stand for derivatives with respect to the proper time of a massive particle or the affine parameter of a massless particle. Particularly, for massive particles of entry velocity $v$, the constants of motion $E$ and $L$ can be written in terms of the impact factor $b$ as:
\begin{equation}\label{EL}
\begin{aligned}
E & =\frac{1}{\sqrt{1-v^2}},\ |L|=\frac{b v}{\sqrt{1-v^2}}=b \sqrt{E^2-\kappa},
\end{aligned}
\end{equation}
with $|L|=b|E|$ also hold for massless particles (e.g. photon). The particle's closest distance $r_0$ to the black hole during its trajectory satisfies the following equation:
\begin{equation}
\left.\dot{r}\right|_{r=r_0}=0,
\end{equation}
which leads to:
\begin{equation}
|L|=\sqrt{\frac{C\left(r_0\right)\left(E^2-\kappa A\left(r_0\right)\right)}{A\left(r_0\right)}}.
\end{equation}
Using eqn. (\ref{EL}), the impact factor $b$ can be computed as:
\begin{equation}
b=\frac{1}{v}\sqrt{\frac{\left[1-\left(1-v^2\right) A(r_0)\right]C(r_0)}{A(r_0)}},
\end{equation}
which returns to $b=\sqrt{\frac{C(r_0)}{A(r_0)}}$ when setting $v=c=1$ (massless particle).

Additionally, let $r_c$ be the radius of ISCO. The following equation will be satisfied if a particle with critical impact parameter $b_c:=\sqrt{\frac{C(r_c)}{A(r_c)}}$ can exactly reach the ISCO:
\begin{equation}\label{eqI}
\left.\frac{\mathrm{d} \dot{r}^2}{\mathrm{~d} r}\right|_{r=r_c}=0.
\end{equation}
the ISCO of both massive and massless particles in arbitrary spacetime with metric described by eqn (\ref{generalG}) can be thus computed by solving eqn. (\ref{eqI}). 

Now, we can discuss the ISCO angle observed by an observer, which can be computed by considering the direction of the test particle's 3-momentum (test particle with critical impact factor $b_c$). Denote the particles 4-momentum at the time it reaches the observer as: $p^\mu=(\dot{t}, \dot{r}, 0, \dot{\phi})$. First, we consider a static observer at $r=D_{\mathrm{OL}}$ and construct a tetrad attached to the observer:
\begin{equation}
\begin{aligned}
e^a_{(\bar{t})} & =\frac{1}{\sqrt{A\left(D_{\mathrm{OL}}\right)}} \left(\frac{\partial}{\partial t}\right)^a,\ e^a_{(\bar{r})} =\frac{1}{\sqrt{B\left(D_{\mathrm{OL}}\right)}} \left(\frac{\partial}{\partial r}\right)^a, \\
e^a_{(\bar{\phi})} & =\frac{1}{\sqrt{C\left(D_{\mathrm{OL}}\right)}} \left(\frac{\partial}{\partial \phi}\right)^a.
\end{aligned}
\end{equation}

Then, the particle’s momentum measured in the observer’s local frame can be computed as:
\begin{equation}
\begin{aligned}
& |p^{\bar{t}}|=\sqrt{A\left(D_{\mathrm{OL}}\right)} |p^t|=\frac{E}{\sqrt{A\left(D_{\mathrm{OL}}\right)}}, \\
& |p^{\bar{r}}|=\sqrt{B\left(D_{\mathrm{OL}}\right)} |p^r|= \sqrt{\frac{E^2}{A\left(D_{\mathrm{OL}}\right)}-\frac{L_c^2}{C\left(D_{\mathrm{OL}}\right)}-\kappa} \\
& |p^{\bar{\phi}}|=\sqrt{C\left(D_{\mathrm{OL}}\right)} |p^\phi|=\frac{L_c}{\sqrt{C\left(D_{\mathrm{OL}}\right)}},
\end{aligned}
\end{equation}
where $L_c:=b_c E$. $|p^{\bar{t}}|$, $|p^{\bar{r}}|$, $|p^{\bar{\theta}}|$, $|p^{\bar{\phi}}|$ are the components of particle's 4-momentum measured in the static observer's local orthonormal frame under the coordinate $\left(\bar{t}, \bar{r}, \bar{\theta}, \bar{\phi}\right)$. The observed angle can thus be computed as:
\begin{equation}\label{thetaf}
\begin{aligned}
    |\theta|&=\arcsin(\frac{|p^{(\phi)}|}{\sqrt{(p^{(r)})^2+(p^{(\phi)})^2}})\\
    &=\arcsin(\frac{|L_c|}{\sqrt{C(D_{\mathrm{OL}})\left(\frac{E^2}{A(D_{\mathrm{OL}})}-\kappa\right)}}),
\end{aligned}
\end{equation}
which is the angle of the ISCO measured by the observer.

To make our computation more physically relevant, we need to further consider observers traveling with an initial velocity. This is because the observations of microscopic black holes are more likely to be particle interactions related experiments. Therefore, we further consider an inertial observer that travels perpendicular to the equatorial plane of the center black hole with a local speed of $v_\mathrm{obs}$ (as observed by a local static observer) at the exact moment it enters the equatorial plane. For such an observer, suppose its 4-velocity moving perpendicular to the equatorial plane at $r=D_{\mathrm{OL}}$ is $u^\mu=\left(u^t, 0, u^\theta, 0\right)$, using $u^\mu u_{\mu}=-1$, this leads to:
\begin{equation}
A\left(D_{\mathrm{OL}}\right)\left(u^t\right)^2-C\left(D_{\mathrm{OL}}\right)\left(u^\theta\right)^2=-1.
\end{equation}
Then, we have the following relation:
\begin{equation}
v_{\mathrm{obs}}=\frac{\mathrm{d} \bar{\theta}}{\mathrm{d} \bar{t}}=\frac{\sqrt{C(D_{\mathrm{OL}})} \mathrm{d} \theta}{\sqrt{A(D_{\mathrm{OL}})} \mathrm{d} t}=\sqrt{\frac{C(D_{\mathrm{OL}})}{A(D_{\mathrm{OL}})}} \cdot \frac{u^\theta}{u^t},
\end{equation}
which gives:
\begin{equation}
u^t=\frac{1}{\sqrt{A\left(D_{\mathrm{OL}}\right)\left(1-v_{\mathrm{obs}}^2\right)}}, \quad u^\theta=\frac{v_{\mathrm{obs}}}{\sqrt{C\left(D_{\mathrm{OL}}\right)\left(1-v_{\mathrm{obs}}^2\right)}} .
\end{equation}
where $d\bar{t}$ and $d\bar{\theta}$ are the proper time and the proper $\bar{\theta}$-direction displacement measured by the local static observer at $r=D_{\mathrm{OL}}$. To compute the deflected angle of the test particle as being observed by the moving observer, we first obtain the components of the tetrad attached to the moving observer with perpendicular $\theta$-direction speed $v_\mathrm{obs}$ as:

\begin{equation}
\begin{aligned}
& e_{(\hat{0})}^\mu=u^{\mu},e_{(\hat{1})}^\mu=\left(0, \frac{1}{\sqrt{B\left(D_{\mathrm{OL}}\right)}}, 0,0\right), \\
& e_{(\hat{2})}^\mu=\left(\frac{v_{\mathrm{obs}}\sqrt{A\left(D_{\mathrm{OL}}\right)}}{\sqrt{1-v_{\mathrm{obs}}^2}}, 0, \frac{\sqrt{C\left(D_{\mathrm{OL}}\right)}}{\sqrt{1-v_{\mathrm{obs}}^2}}, 0\right),\\
& e_{(\overline{3})}^\mu=\left(0,0,0, \frac{1}{\sqrt{C\left(D_{\mathrm{OL}}\right)}}\right) .
\end{aligned}
\end{equation}
This leads to the particle's momentum measured by the moving observer:
\begin{equation}
\begin{aligned}
& p^{(\hat{t})}=\frac{E}{\sqrt{A\left(D_{\mathrm{OL}}\right)\left(1-v_{\mathrm{obs}}^2\right)}}, p^{(\hat{r})}= |p^{(\bar{r})}|,\\
& p^{(\hat{\theta})}=\frac{v_{\mathrm{obs}} E}{\sqrt{1-v_{\mathrm{obs}}^2}}, p^{(\hat{\phi})}= |p^{(\bar{\phi})}|.
\end{aligned}
\end{equation}
Finally, the observed angle is computed as:
\begin{widetext}
\begin{equation}\label{thetafa}
    |\theta|=\arcsin\left(\sqrt{\frac{(p^{(\hat{\phi})})^2+p^{(\hat{\theta})})^2}{(p^{(\hat{r})})^2+(p^{(\hat{\phi})})^2+p^{(\hat{\theta})})^2}}\right)=\arcsin \left(\sqrt{\frac{A\left(D_{\mathrm{OL}}\right)v_{\mathrm{obs}}^2 E^2+\frac{L^2\left(1-v_{\mathrm{obs}}^2\right)}{C\left(D_{\mathrm{OL}}\right)}}{(1-v_{\mathrm{obs}}^2)\left(\frac{E^2}{A\left(D_{\mathrm{OL}}\right)}-\kappa\right)+A\left(D_{\mathrm{OL}}\right)v_{\mathrm{obs}}^2 E^2}}\right).
\end{equation}
\end{widetext}
As $v_\mathrm{obs}=0$, this result returns to eqn. (\ref{thetaf}).

\begin{figure*}
\includegraphics[height=3.6cm]{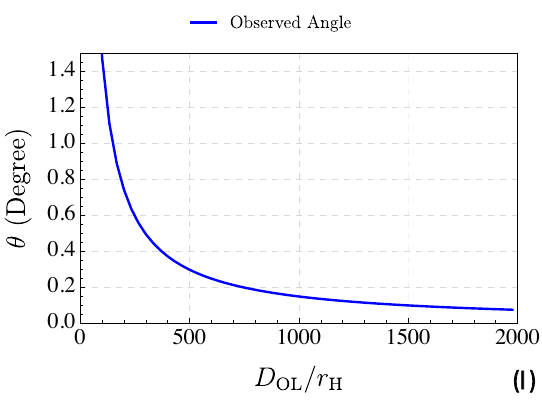} 
\includegraphics[height=3.6cm]{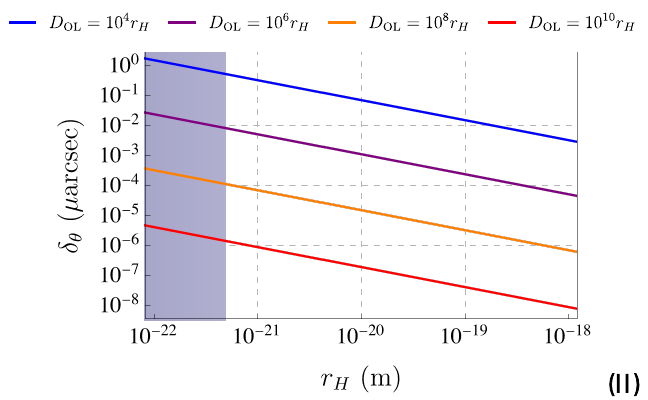} 
\includegraphics[height=3.6cm]{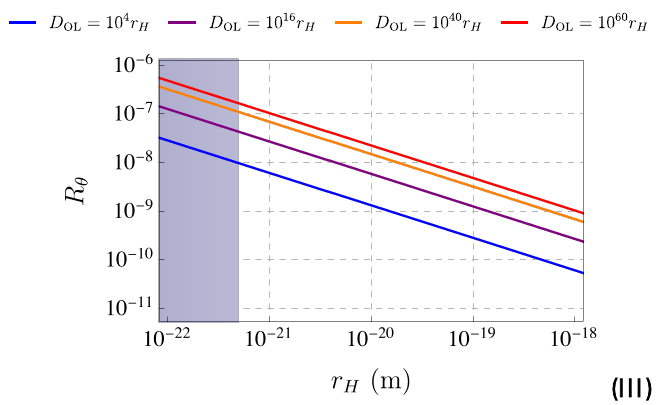} 
\includegraphics[height=3.6cm]{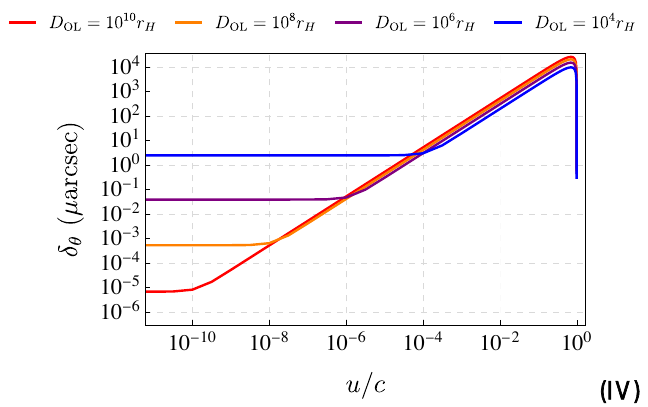} 
\includegraphics[height=3.6cm]{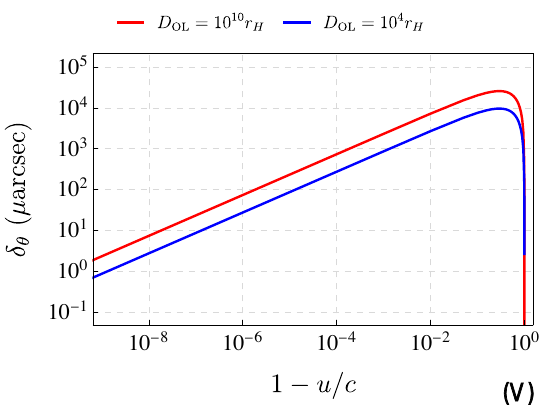} 
\includegraphics[height=3.6cm]{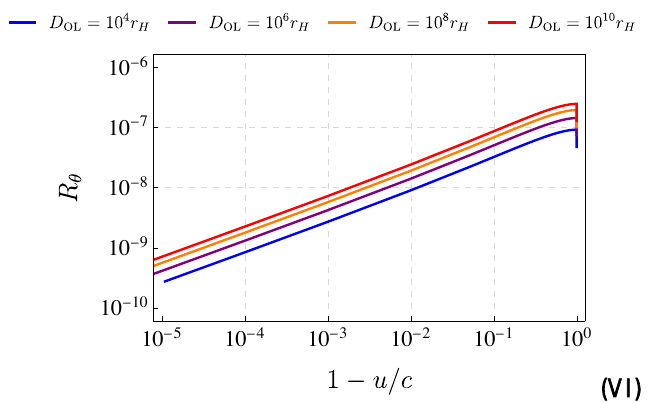} 
\includegraphics[height=4cm]{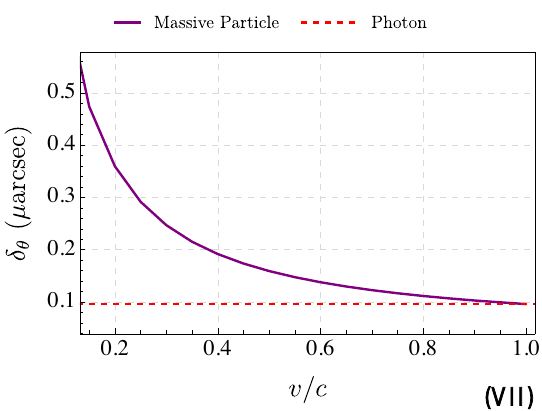} 
\includegraphics[height=4cm]{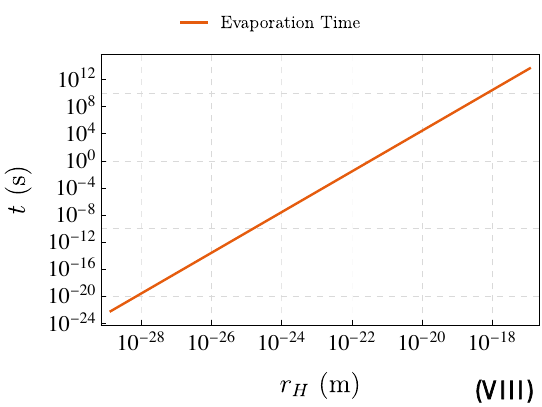} 
\includegraphics[height=4cm]{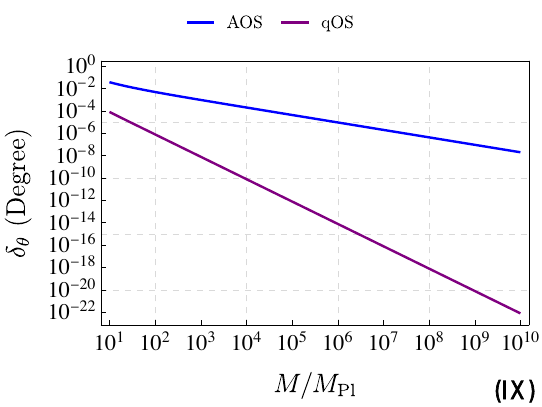} 
\caption{\label{fig:1} The main results of our work: (I) the observed angle of photon ISCO for both AOS black hole and Schwarzschild black hole. (II) The absolute difference of the observed angle of photon ISCO at scale $10^{-22}m \leq r_H\leq 10^{-18}m$ for observers at four different distances to the center black hole, the purple-grey area represents the deep quantum region of the largest energy Gamma ray currently observed. (III) The relative difference of the observed angle of photon ISCO at scale $10^{-22}m \leq r_H\leq 10^{-18}m$. (IV) The absolute difference of the observed angle of photon ISCO for $r_H=10^{-22}m$ when considering the aberration effect in eqn. (\ref{thetafa}), in the limit $v_{obs}\rightarrow 0$. (V) The absolute difference of the observed angle of photon ISCO for $r_H=10^{-22}m$ when considering the aberration effect in eqn. (\ref{thetafa}), in the limit $v_{obs}\rightarrow c$. (VI) The relative difference of the observed angle of photon ISCO for $r_H=10^{-22}m$ when considering the aberration effect in eqn. (\ref{thetafa}), in the limit $v_{obs}\rightarrow c$. (VII) Observed angle of ISCO of massive particle with velocity $v$ (relative to the angle of photon ISCO) for $r_H=10^{-22}m$, $D_{\mathrm{OL}}=10^4 r_H$. (VIII) Hawking evaporation time of black holes of different sizes. (IX) The quantum correction of the AOS model versus the quantum Oppenheimer model as the size of the microscopic black hole approaches the Planck scale.}
\end{figure*}

\section{Results}\label{sec3}

In this work, we compute the quantum corrections to the ISCO induced by a microscopic AOS black hole, where the quantum parameter is obtained authentically from the original model. Recall that in \citep{Li:2024afr}, we have computed the quantum correction of the measured angle $\theta_{\infty}$ of the ISCO of $\mathrm{SgrA}^*$ observed from Earth. The quantum corrections are in the order of $10^{-31}\mu \mathrm{arcsec}$. This is practically unobservable due to other potential corrections at much larger scale, such as black hole rotation and electric charge. 

This situation may change by considering eqn. (\ref{AOS}) and eqn. (\ref{lpm}). As the black hole mass $M$ decreases (corresponding to a microscopic black hole when $M$ becomes extremely small), the quantum correction from the LQG black hole increases. The main results are shown in Fig. \ref{fig:1}. First, we introduce the following notations for the absolute difference $\delta_\theta$ and relative difference $R_\theta$:
\begin{equation}
    \delta_{\theta}:=|\theta_{LQG}-\theta_{0}|,\ R_{\theta}:=\frac{\delta_{\theta}}{\theta_{0}}.
\end{equation}
where $\theta_{LQG}$ is the observed angle of ISCO for AOS black hole, $\theta_{0}$ is the observed angle of ISCO for Schwarzschild black hole, $\delta_{\theta}$ is the absolute difference, while $R_{\theta}$ is the difference relative to $\theta_{0}$.

Fig. \ref{fig:1} (I) shows the observed angle of photon ISCO as seen by an observer at a distance $D_{\mathrm{OL}}$ from the center of the black hole. Since the quantum correction is generally very small, all computed results correspond to the same curve of $\theta$.

In Fig. \ref{fig:1} (II), $\delta_{\theta}$ of photon is computed with $D_{\mathrm{OL}}$ being set as $10^4 r_H$ (blue), $10^6 r_H$ (purple), $10^8 r_H$ (orange), $10^{10} r_H$ (red), respectively. The region painted in purple-grey represents the deep quantum region where the radius of the black hole horizon becomes smaller than the wavelength of the highest energy gamma-ray known to exist ( $2.5 \mathrm{PeV}$ \citep{LHAASO:2023uhj}). As we can see from this graph, the quantum correction is, in fact, larger for the observer closer to the black hole. The observer with distance $10^4 r_H$ can observe a difference of roughly $0.5 \mu\mathrm{arcsec}$ for $r_H=5\times 10^{-22}m$. for $D_{\mathrm{OL}}=10^{10} r_H$, the absolute difference decrease to $10^{-6} \mu\mathrm{arcsec}$, which is still $10^{24}$ larger than the previous results for macroscopic black hole gravitational lenses.

Fig. \ref{fig:1} (III) shows the relative difference $R_{\theta}:=\frac{|\theta_{\mathrm{LQG}}-\theta_{\mathrm{0}}|}{\theta_{\mathrm{0}}}$ between the observed angle $\theta_{\mathrm{LQG}}$ from LQG black hole and $\theta_0$ from Schwarzschild black hole. The overall relative difference is larger than $10^{-10}$, suggesting that ideally, a noticeable difference could be observed between the AOS black hole and the Schwarzschild black hole at this scale. Here, it is worth noting that the relative difference becomes larger as $D_{\mathrm{OL}}$ increases, which is counter-intuitive for an asymptotically flat black hole considering eqn. (\ref{thetaf}). The reason for this to happen is because of the presence of $\left(\frac{r}{r_H}\right)^{2\epsilon}$ in $A_{AOS}$, which is slowly divergent as $r\rightarrow\infty$. The implication of this slow divergence has been analyzed in detail in \citep{Ashtekar:2020ckv}. In Fig. \ref{fig:1} (III) we can easily see that, as the distance $D_{OL}$ increases, the increase in relative difference quickly converges to the red line within the scale of the entire observable universe, thus the resulting increase in quantum correction at very large $D_{\mathrm{OL}}$ due to $\left(\frac{r}{r_H}\right)^{2\epsilon}$ is negligible.

Here, we should further discuss the possible observation error created by the observer's motion during the observation period. First, note that the estimation of this error depends heavily on the specific experimental setup, which can be very complicated and is beyond the main topic of this work. Since the observer is far from the black hole, a simple estimation can be performed by considering the classical Newtonian laws of motion. For $r_H=10^{-21}m$, $D_{\mathrm{OL}}=10^8r_H$, the acceleration of the observer can be approximated by (when $t$ is extremely small) $a=\frac{G M}{D_{\mathrm{OL}}^2}=4.5\times 10^{21} m/s^2$. If we estimate the total time of observation as the time in which photons travel from the vicinity of the black hole to the observer (in fact, the actual observation process can take much less time), then $t=\frac{D_{\mathrm{OL}}}{c}=3.3\times 10^{-22}s$. It can be estimated that:
\begin{equation*}
    \begin{aligned}
        &v=a\cdot t=1.5m/s,\ \beta:=\frac{v^2}{c^2}=1.7\times 10^{-17},\\
        &\delta x=\frac{1}{2}a\cdot t^2=2.5\times10^{-22}m=2.5\times 10^{-9}D_{OL}.
    \end{aligned}
\end{equation*}
By comparing $\delta x$ and $\beta$ with the results in Fig. \ref{fig:1} (III), it can be seen that it is possible for the error estimation of observer motion to be lower than the relative difference $R_{\theta}$ induced by the quantum correction.

We further discuss the motion aberration effect described by eqn. (\ref{thetafa}) in Fig. \ref{fig:1} (IV)-(VI), by considering a moving observer with a 3-velocity $v_\mathrm{obs}$ in the $\theta$ direction (perpendicular to the equatorial plane). Fig. \ref{fig:1} (IV) and Fig. \ref{fig:1} (V) show the absolute difference of the deflection angle when $v_\mathrm{obs}\rightarrow 0$ and $v_\mathrm{obs}\rightarrow c$, respectively. In Fig. \ref{fig:1} (IV), it is shown that at first, when $v_\mathrm{obs}$ is very small, $\delta\theta$ is almost unchanged from the $v_\mathrm{obs}=0$ case. Then $\delta\theta$ begins to increase with the same slope in the log-log graph, regardless of different $D_\mathrm{OL}$. Finally, a sharp decrease occurs as $v_\mathrm{obs}/c\rightarrow1$. The reason for these effects to happen is because: First, $\delta\theta$ is non-zero when $v_\mathrm{obs}=0$. At very small $v_\mathrm{obs}$, the aberration effect is much smaller than $\delta\theta|_{v_\mathrm{obs}=0}$, which results in the flat curves at the very left of Fig. \ref{fig:1} (IV). Also, the same slope in the log-log graph for different choices of $D_\mathrm{OL}$ in $10^{-4}\leq u/c\leq 10^0$ indicates that the impact of $D_\mathrm{OL}$ is minimal in this range. This mainly comes from the combination of large $C(D_\mathrm{OL})$ and $A(D_\mathrm{OL})\rightarrow 1$ in all of the choices. Fig. \ref{fig:1} (V) shows the region of sharply decrease in Fig. \ref{fig:1} (IV) when $v_\mathrm{obs}/c=1$. As $v_\mathrm{obs}\rightarrow c$, the decrease of $\delta\theta$ is proportional to $\sqrt{1-u/c}$. Fig. \ref{fig:1} (VI) shows the relative difference $R_{\theta}$ when $v_\mathrm{obs}\rightarrow c$, the trend is similar in general to Fig. \ref{fig:1} (V), and the relative difference is still greater than $10^{-10}$ even when $1-v_\mathrm{obs}/c<10^{-5}$, namely when the observer's 3-velocity, as being measured by a local static observer at $r=D_\mathrm{OL}$, is extremely close to the speed of light. Overall, we can conclude from Fig. \ref{fig:1} (IV)-(VI) that the motion aberration effect in general does not drastically change the result, except when the observer travels at extremely high velocity ($1-v_\mathrm{obs}/c<10^{-5}$).

Fig. \ref{fig:1} (VII) shows the corrections $\delta_{\theta}$ of the observed angle of ISCO for massive particles relative to the corrections to the photon's observed angle $\delta_{\theta,\mathrm{Photon}}$. It approaches the photon results as $ v\rightarrow 1$. Also, the correction is much larger for particles with smaller $v$. These particles can serve as additional choices for the test particle.

In Fig. (\ref{fig:1}) (VIII), the Hawking evaporation time $\tau=\frac{5120 \pi G^2 M^3}{\hbar c^4}$ of the center black hole is plotted. As can be seen from this graph, the Hawking evaporation remains slow at $r_H=10^{-22} m$, when compared to the time duration for the observation $\tau=3.3\times 10^{-23}s$. This ensures the stability of the central black hole during the observation period.

We also compare the quantum corrections of the AOS model and the quantum Oppenheimer-Schneider (qOS) model \citep{Lewandowski:2022zce,Zhang2024c}:
\begin{equation*}
\begin{aligned}
& A_{qOS}(r)=1-\frac{2M}{r}+\frac{\zeta_{qOS} M^2}{r^4}, \\
& B_{qOS}(r)=A_{qOS}(r)^{-1}, \\
& C_{qOS}(r)=r^2,
\end{aligned}
\end{equation*}
where $\zeta_{qOS}=16\sqrt{3}\gamma^3l^2_p$. The gravitational lensing effect of this model has also been studied previously in \citep{Zhao:2024elr}, considering $\zeta_{qOS}\sim 1$. In this work, we keep the quantum parameter as $\zeta_{qOS}=16\sqrt{3}\gamma^3l^2_p$. For the qOS model, previous computation shows that its quantum correction is extremely small at large scale ($\sim 10^{-90} \mu \mathrm{arcsec}$), which is mainly due to the fact that its difference from the Schwarzschild black hole is caused by a combination of $1/r^4$ and $l_p^2$, which are both tiny quantities at large scale. The situation changes drastically, however, as the size of the black hole approaches the Planck scale. Under such an extreme, $1/r^4$ becomes very large and thus fully compensates for the small quantum parameter. Fig. (\ref{fig:1}) (VI) shows the absolute difference $\delta_{\theta}$ as the mass of the black hole becomes close to the Planck mass $M_{\mathrm{Pl}}=2.176\times 10^{-8}\ \mathrm{kg}$. As we can see from Fig. (\ref{fig:1}) (VI), the quantum corrections of both the AOS and qOS model increases significantly at small $M/ M_{\mathrm{Pl}}$. However, the quantum correction of the qOS model is still smaller than the AOS model. Also, the slope for the Log-Log plot is much steeper for the qOS model, thus explaining its extremely small quantum correction at large scale.


\section{Discussion}\label{sec4}
In summary, there are four key points we would like to make regarding the obtained results:
\begin{itemize}
    \item Based on our study in this work, we discovered that as the black hole size decreases to the microscopic scale, there is a significant increase in the magnitude of quantum correction in terms of the observed angle of ISCO of test particles. Noticeably, at the scale $r_H \sim 10^{-20}\ m$, the relative difference between the Schwarzschild results and the quantum corrected black hole can be higher than $10^{-10}$, and higher level quantum corrections can be observed when considering low velocity massive particles. Meanwhile, the absolute difference can reach $1 \mu \mathrm{arcsec}$. Notably, we also considered several fundamental constraints, including the wavelength of the test particles, black hole evaporation speed, and an estimation of observer motion within the observation time window. So far, we have found no fundamental constraint that prohibits the observation at this scale. We also confirm previous results in LQG about the large quantum correction near the Planck scale. Moreover, we compare the correction of AOS black hole and the quantum Openheimer-Snyder model, whose quantum correction at large scale is much smaller than the AOS black hole. It turns out that the quantum correction generated by the qOS model gets larger rapidly as the black hole approaches the Planck scale.

    \item Since we are discussing the observable results from microscopic black holes that are extremely small, it is important to consider the real-world physical scenario in which our calculations apply: First, it should be noted that there is only one black hole in the experiment. However, given a microscopic black hole whose mass is precisely measured, it is theoretically possible to determine whether the ISCO of particles around the black hole follows the computed result for the Schwarzschild black hole or the loop quantum black hole, thus providing potential evidence for quantum gravity. Second, unlike massive black holes that can be directly observed and measured in astronomy, microscopic black holes with a horizon radius of $10^{-20} m$ are unlikely to be observed with sufficient precision to test the effect of quantum gravity in any astronomical observation. However, our current understanding of physics does not rule out the possibility of artificially creating such black holes. An artificially created microscopic black hole can be ideal for such an observation because the small relative difference requires high precision in measuring the black hole mass and the deflection angle. An artificially created black hole might be controllable regarding its exact mass and other factors, such as black hole rotation, charge, etc. (Ideally, these factors should be as small as possible for the highest precision to be achieved). Hypothetically, such a microscopic black hole could be manufactured in a two-step process, i.e., by first establishing a black hole core using a small amount of extremely high-energy materials and then injecting regular matter to enlarge it to the desired size. Third, in this work, we only considered the particle deflection in the equatorial plane. Although such calculation may share some insight into how large the quantum correction from LQG black holes might be, in reality the actual physical process is much more complicated: If we take a beam of particles (passing through the vicinity region of the microscopic black hole) as the observer to measure how much the deflected test particles affect its trajectory through particle scattering, not only could it be affected by the deflected test particles in the equatorial plane, but it also can be affected by test particles throughout its entire trajectory. Therefore, we do not yet know how to actually measure the test particle's ISCO in this scenario. There is a lot of extra work to be done before an actual experiment that can exclusively capture the deflection of test particles at such a microscopic scale can be designed.

    \item Since the source of the increase of quantum correction of ISCO comes directly from the quantum corrections to the metric, our results can be expected to apply to other gravitational lensing observables as well. One apparent exception, however, is the time delay of the test particle following different deflected paths. This is because the time for a relativistic particle to traverse a microscopic black hole is almost instantaneous, making any quantum correction impossible to detect. Our work studies the quantum correction of ISCO by LQG black hole, yet our conclusion can also be generalized for other black hole models of quantum gravity, e.g., quantum black holes constructed from effective field theory \cite{Donoghue:2001qc,Kirilin:2006en,Battista:2023iyu,Wang:2025fmz} and black hole models in string theory \cite{Maldacena:1996ky,Bena:2022rna}, etc.
    
    \item Although the quantum correction is still small in the case of microscopic black holes, based on the analysis above, we have found no fundamental physical constraint to prohibit the detection of quantum gravity effects at such a scale. Suppose the correction from models of loop quantum black hole or other black hole models from quantum gravity can further increase by several (e.g., $10^5$) orders of magnitude. In that case, the constraints discussed in this work will be significantly easier to overcome. Here we suggest that, at least for LQBH models, it is still possible to make the effect larger. There are two main directions to consider: First is the theoretical advancement of LQG, such as implementing effective dynamics from the full theory of LQG and resolving existing regularization ambiguities in the theory. These theoretical advancements may provide new sources of quantum corrections in LQG. Second is further considering the quantum properties of the test particle during its propagation near the black hole using quantum field theory of curved space-time, which can be used to evaluate the quantum corrections more accurately. However, one of the important dilemmas in this approach is the actual computation of off-center particle propagation, which may require the quantization of particle states with some asymmetrical characteristics. 

\end{itemize}

\section*{Acknowledgements}
The authors thank Hongguang Liu for the helpful discussion. This work is supported by National Natural Science Foundation of China (NSFC) with Grants No.12275087.

\providecommand{\noopsort}[1]{}\providecommand{\singleletter}[1]{#1}%


\begin{thebibliography}{42}%
\makeatletter
\providecommand \@ifxundefined [1]{%
 \@ifx{#1\undefined}
}%
\providecommand \@ifnum [1]{%
 \ifnum #1\expandafter \@firstoftwo
 \else \expandafter \@secondoftwo
 \fi
}%
\providecommand \@ifx [1]{%
 \ifx #1\expandafter \@firstoftwo
 \else \expandafter \@secondoftwo
 \fi
}%
\providecommand \natexlab [1]{#1}%
\providecommand \enquote  [1]{``#1''}%
\providecommand \bibnamefont  [1]{#1}%
\providecommand \bibfnamefont [1]{#1}%
\providecommand \citenamefont [1]{#1}%
\providecommand \href@noop [0]{\@secondoftwo}%
\providecommand \href [0]{\begingroup \@sanitize@url \@href}%
\providecommand \@href[1]{\@@startlink{#1}\@@href}%
\providecommand \@@href[1]{\endgroup#1\@@endlink}%
\providecommand \@sanitize@url [0]{\catcode `\\12\catcode `\$12\catcode
  `\&12\catcode `\#12\catcode `\^12\catcode `\_12\catcode `\%12\relax}%
\providecommand \@@startlink[1]{}%
\providecommand \@@endlink[0]{}%
\providecommand \url  [0]{\begingroup\@sanitize@url \@url }%
\providecommand \@url [1]{\endgroup\@href {#1}{\urlprefix }}%
\providecommand \urlprefix  [0]{URL }%
\providecommand \Eprint [0]{\href }%
\providecommand \doibase [0]{https://doi.org/}%
\providecommand \selectlanguage [0]{\@gobble}%
\providecommand \bibinfo  [0]{\@secondoftwo}%
\providecommand \bibfield  [0]{\@secondoftwo}%
\providecommand \translation [1]{[#1]}%
\providecommand \BibitemOpen [0]{}%
\providecommand \bibitemStop [0]{}%
\providecommand \bibitemNoStop [0]{.\EOS\space}%
\providecommand \EOS [0]{\spacefactor3000\relax}%
\providecommand \BibitemShut  [1]{\csname bibitem#1\endcsname}%
\let\auto@bib@innerbib\@empty
\bibitem [{\citenamefont {Thiemann}(2007)}]{Thiemann:2007pyv}%
  \BibitemOpen
  \bibfield  {author} {\bibinfo {author} {\bibfnamefont {T.}~\bibnamefont
  {Thiemann}},\ }\href {https://doi.org/10.1017/CBO9780511755682} {\emph
  {\bibinfo {title} {{Modern Canonical Quantum General Relativity}}}},\
  Cambridge Monographs on Mathematical Physics\ (\bibinfo  {publisher}
  {Cambridge University Press},\ \bibinfo {year} {2007})\BibitemShut {NoStop}%
\bibitem [{\citenamefont {Rovelli}\ and\ \citenamefont
  {Vidotto}(2014)}]{Rovelli:2014ssa}%
  \BibitemOpen
  \bibfield  {author} {\bibinfo {author} {\bibfnamefont {C.}~\bibnamefont
  {Rovelli}}\ and\ \bibinfo {author} {\bibfnamefont {F.}~\bibnamefont
  {Vidotto}},\ }\href@noop {} {\emph {\bibinfo {title} {{Covariant Loop Quantum
  Gravity}: {An Elementary Introduction to Quantum Gravity and Spinfoam
  Theory}}}},\ Cambridge Monographs on Mathematical Physics\ (\bibinfo
  {publisher} {Cambridge University Press},\ \bibinfo {year}
  {2014})\BibitemShut {NoStop}%
\bibitem [{\citenamefont {Ashtekar}\ and\ \citenamefont
  {Pullin}(2017)}]{Ashtekar:2017yom}%
  \BibitemOpen
  \bibinfo {editor} {\bibfnamefont {A.}~\bibnamefont {Ashtekar}}\ and\ \bibinfo
  {editor} {\bibfnamefont {J.}~\bibnamefont {Pullin}},\ eds.,\ \href
  {https://doi.org/10.1142/10445} {\emph {\bibinfo {title} {{Loop Quantum
  Gravity}: {The First 30 Years}}}},\ \bibinfo {series} {100 Years of General
  Relativity}, Vol.~\bibinfo {volume} {4}\ (\bibinfo  {publisher} {World
  Scientific},\ \bibinfo {year} {2017})\BibitemShut {NoStop}%
\bibitem [{\citenamefont {Ashtekar}\ and\ \citenamefont
  {Singh}(2011)}]{Ashtekar:2011ni}%
  \BibitemOpen
  \bibfield  {author} {\bibinfo {author} {\bibfnamefont {A.}~\bibnamefont
  {Ashtekar}}\ and\ \bibinfo {author} {\bibfnamefont {P.}~\bibnamefont
  {Singh}},\ }\bibfield  {title} {\bibinfo {title} {{Loop Quantum Cosmology: A
  Status Report}},\ }\href {https://doi.org/10.1088/0264-9381/28/21/213001}
  {\bibfield  {journal} {\bibinfo  {journal} {Class. Quant. Grav.}\ }\textbf
  {\bibinfo {volume} {28}},\ \bibinfo {pages} {213001} (\bibinfo {year}
  {2011})},\ \Eprint {https://arxiv.org/abs/1108.0893} {arXiv:1108.0893
  [gr-qc]} \BibitemShut {NoStop}%
\bibitem [{\citenamefont {Modesto}(2006)}]{Modesto:2005zm}%
  \BibitemOpen
  \bibfield  {author} {\bibinfo {author} {\bibfnamefont {L.}~\bibnamefont
  {Modesto}},\ }\bibfield  {title} {\bibinfo {title} {{Loop quantum black
  hole}},\ }\href {https://doi.org/10.1088/0264-9381/23/18/006} {\bibfield
  {journal} {\bibinfo  {journal} {Class. Quant. Grav.}\ }\textbf {\bibinfo
  {volume} {23}},\ \bibinfo {pages} {5587} (\bibinfo {year} {2006})},\ \Eprint
  {https://arxiv.org/abs/gr-qc/0509078} {arXiv:gr-qc/0509078} \BibitemShut
  {NoStop}%
\bibitem [{\citenamefont {Ashtekar}\ \emph
  {et~al.}(2018{\natexlab{a}})\citenamefont {Ashtekar}, \citenamefont
  {Olmedo},\ and\ \citenamefont {Singh}}]{Ashtekar:2018lag}%
  \BibitemOpen
  \bibfield  {author} {\bibinfo {author} {\bibfnamefont {A.}~\bibnamefont
  {Ashtekar}}, \bibinfo {author} {\bibfnamefont {J.}~\bibnamefont {Olmedo}},\
  and\ \bibinfo {author} {\bibfnamefont {P.}~\bibnamefont {Singh}},\ }\bibfield
   {title} {\bibinfo {title} {{Quantum Transfiguration of Kruskal Black
  Holes}},\ }\href {https://doi.org/10.1103/PhysRevLett.121.241301} {\bibfield
  {journal} {\bibinfo  {journal} {Phys. Rev. Lett.}\ }\textbf {\bibinfo
  {volume} {121}},\ \bibinfo {pages} {241301} (\bibinfo {year}
  {2018}{\natexlab{a}})},\ \Eprint {https://arxiv.org/abs/1806.00648}
  {arXiv:1806.00648 [gr-qc]} \BibitemShut {NoStop}%
\bibitem [{\citenamefont {Gambini}\ \emph {et~al.}(2020)\citenamefont
  {Gambini}, \citenamefont {Olmedo},\ and\ \citenamefont
  {Pullin}}]{Gambini:2020nsf}%
  \BibitemOpen
  \bibfield  {author} {\bibinfo {author} {\bibfnamefont {R.}~\bibnamefont
  {Gambini}}, \bibinfo {author} {\bibfnamefont {J.}~\bibnamefont {Olmedo}},\
  and\ \bibinfo {author} {\bibfnamefont {J.}~\bibnamefont {Pullin}},\
  }\bibfield  {title} {\bibinfo {title} {{Spherically symmetric loop quantum
  gravity: analysis of improved dynamics}},\ }\href
  {https://doi.org/10.1088/1361-6382/aba842} {\bibfield  {journal} {\bibinfo
  {journal} {Class. Quant. Grav.}\ }\textbf {\bibinfo {volume} {37}},\ \bibinfo
  {pages} {205012} (\bibinfo {year} {2020})},\ \Eprint
  {https://arxiv.org/abs/2006.01513} {arXiv:2006.01513 [gr-qc]} \BibitemShut
  {NoStop}%
\bibitem [{\citenamefont {Kelly}\ \emph {et~al.}(2020)\citenamefont {Kelly},
  \citenamefont {Santacruz},\ and\ \citenamefont
  {Wilson-Ewing}}]{Kelly:2020uwj}%
  \BibitemOpen
  \bibfield  {author} {\bibinfo {author} {\bibfnamefont {J.~G.}\ \bibnamefont
  {Kelly}}, \bibinfo {author} {\bibfnamefont {R.}~\bibnamefont {Santacruz}},\
  and\ \bibinfo {author} {\bibfnamefont {E.}~\bibnamefont {Wilson-Ewing}},\
  }\bibfield  {title} {\bibinfo {title} {{Effective loop quantum gravity
  framework for vacuum spherically symmetric spacetimes}},\ }\href
  {https://doi.org/10.1103/PhysRevD.102.106024} {\bibfield  {journal} {\bibinfo
   {journal} {Phys. Rev. D}\ }\textbf {\bibinfo {volume} {102}},\ \bibinfo
  {pages} {106024} (\bibinfo {year} {2020})},\ \Eprint
  {https://arxiv.org/abs/2006.09302} {arXiv:2006.09302 [gr-qc]} \BibitemShut
  {NoStop}%
\bibitem [{\citenamefont {Han}\ and\ \citenamefont {Liu}(2022)}]{Han:2020uhb}%
  \BibitemOpen
  \bibfield  {author} {\bibinfo {author} {\bibfnamefont {M.}~\bibnamefont
  {Han}}\ and\ \bibinfo {author} {\bibfnamefont {H.}~\bibnamefont {Liu}},\
  }\bibfield  {title} {\bibinfo {title} {{Improved effective dynamics of
  loop-quantum-gravity black hole and Nariai limit}},\ }\href
  {https://doi.org/10.1088/1361-6382/ac44a0} {\bibfield  {journal} {\bibinfo
  {journal} {Class. Quant. Grav.}\ }\textbf {\bibinfo {volume} {39}},\ \bibinfo
  {pages} {035011} (\bibinfo {year} {2022})},\ \Eprint
  {https://arxiv.org/abs/2012.05729} {arXiv:2012.05729 [gr-qc]} \BibitemShut
  {NoStop}%
\bibitem [{\citenamefont {Lewandowski}\ \emph {et~al.}(2023)\citenamefont
  {Lewandowski}, \citenamefont {Ma}, \citenamefont {Yang},\ and\ \citenamefont
  {Zhang}}]{Lewandowski:2022zce}%
  \BibitemOpen
  \bibfield  {author} {\bibinfo {author} {\bibfnamefont {J.}~\bibnamefont
  {Lewandowski}}, \bibinfo {author} {\bibfnamefont {Y.}~\bibnamefont {Ma}},
  \bibinfo {author} {\bibfnamefont {J.}~\bibnamefont {Yang}},\ and\ \bibinfo
  {author} {\bibfnamefont {C.}~\bibnamefont {Zhang}},\ }\bibfield  {title}
  {\bibinfo {title} {{Quantum Oppenheimer-Snyder and Swiss Cheese Models}},\
  }\href {https://doi.org/10.1103/PhysRevLett.130.101501} {\bibfield  {journal}
  {\bibinfo  {journal} {Phys. Rev. Lett.}\ }\textbf {\bibinfo {volume} {130}},\
  \bibinfo {pages} {101501} (\bibinfo {year} {2023})},\ \Eprint
  {https://arxiv.org/abs/2210.02253} {arXiv:2210.02253 [gr-qc]} \BibitemShut
  {NoStop}%
\bibitem [{\citenamefont {Zhang}(2023)}]{Zhang:2023yps}%
  \BibitemOpen
  \bibfield  {author} {\bibinfo {author} {\bibfnamefont {X.}~\bibnamefont
  {Zhang}},\ }\bibfield  {title} {\bibinfo {title} {{Loop Quantum Black
  Hole}},\ }\href {https://doi.org/10.3390/universe9070313} {\bibfield
  {journal} {\bibinfo  {journal} {Universe}\ }\textbf {\bibinfo {volume} {9}},\
  \bibinfo {pages} {313} (\bibinfo {year} {2023})},\ \Eprint
  {https://arxiv.org/abs/2308.10184} {arXiv:2308.10184 [gr-qc]} \BibitemShut
  {NoStop}%
\bibitem [{\citenamefont {Giesel}\ \emph {et~al.}(2023)\citenamefont {Giesel},
  \citenamefont {Liu}, \citenamefont {Rullit}, \citenamefont {Singh},\ and\
  \citenamefont {Weigl}}]{Giesel:2023tsj}%
  \BibitemOpen
  \bibfield  {author} {\bibinfo {author} {\bibfnamefont {K.}~\bibnamefont
  {Giesel}}, \bibinfo {author} {\bibfnamefont {H.}~\bibnamefont {Liu}},
  \bibinfo {author} {\bibfnamefont {E.}~\bibnamefont {Rullit}}, \bibinfo
  {author} {\bibfnamefont {P.}~\bibnamefont {Singh}},\ and\ \bibinfo {author}
  {\bibfnamefont {S.~A.}\ \bibnamefont {Weigl}},\ }\bibfield  {title} {\bibinfo
  {title} {{Embedding generalized LTB models in polymerized spherically
  symmetric spacetimes}},\ }\href@noop {} {\  (\bibinfo {year} {2023})},\
  \Eprint {https://arxiv.org/abs/2308.10949} {arXiv:2308.10949 [gr-qc]}
  \BibitemShut {NoStop}%
\bibitem [{\citenamefont {Lin}\ and\ \citenamefont
  {Zhang}(2024)}]{Lin:2024flv}%
  \BibitemOpen
  \bibfield  {author} {\bibinfo {author} {\bibfnamefont {J.}~\bibnamefont
  {Lin}}\ and\ \bibinfo {author} {\bibfnamefont {X.}~\bibnamefont {Zhang}},\
  }\bibfield  {title} {\bibinfo {title} {{Effective four-dimensional loop
  quantum black hole with a cosmological constant}},\ }\href
  {https://doi.org/10.1103/PhysRevD.110.026002} {\bibfield  {journal} {\bibinfo
   {journal} {Phys. Rev. D}\ }\textbf {\bibinfo {volume} {110}},\ \bibinfo
  {pages} {026002} (\bibinfo {year} {2024})},\ \Eprint
  {https://arxiv.org/abs/2402.13638} {arXiv:2402.13638 [gr-qc]} \BibitemShut
  {NoStop}%
\bibitem [{\citenamefont {Ashtekar}\ \emph {et~al.}(2006)\citenamefont
  {Ashtekar}, \citenamefont {Pawlowski},\ and\ \citenamefont
  {Singh}}]{Ashtekar:2006rx}%
  \BibitemOpen
  \bibfield  {author} {\bibinfo {author} {\bibfnamefont {A.}~\bibnamefont
  {Ashtekar}}, \bibinfo {author} {\bibfnamefont {T.}~\bibnamefont
  {Pawlowski}},\ and\ \bibinfo {author} {\bibfnamefont {P.}~\bibnamefont
  {Singh}},\ }\bibfield  {title} {\bibinfo {title} {{Quantum nature of the big
  bang}},\ }\href {https://doi.org/10.1103/PhysRevLett.96.141301} {\bibfield
  {journal} {\bibinfo  {journal} {Phys. Rev. Lett.}\ }\textbf {\bibinfo
  {volume} {96}},\ \bibinfo {pages} {141301} (\bibinfo {year} {2006})},\
  \Eprint {https://arxiv.org/abs/gr-qc/0602086} {arXiv:gr-qc/0602086}
  \BibitemShut {NoStop}%
\bibitem [{\citenamefont {Wilson-Ewing}(2017)}]{Wilson-Ewing:2016yan}%
  \BibitemOpen
  \bibfield  {author} {\bibinfo {author} {\bibfnamefont {E.}~\bibnamefont
  {Wilson-Ewing}},\ }\bibfield  {title} {\bibinfo {title} {{Testing loop
  quantum cosmology}},\ }\href {https://doi.org/10.1016/j.crhy.2017.02.004}
  {\bibfield  {journal} {\bibinfo  {journal} {Comptes Rendus Physique}\
  }\textbf {\bibinfo {volume} {18}},\ \bibinfo {pages} {207} (\bibinfo {year}
  {2017})},\ \Eprint {https://arxiv.org/abs/1612.04551} {arXiv:1612.04551
  [gr-qc]} \BibitemShut {NoStop}%
\bibitem [{\citenamefont {Li}\ \emph {et~al.}(2021)\citenamefont {Li},
  \citenamefont {Singh},\ and\ \citenamefont {Wang}}]{Li:2021mop}%
  \BibitemOpen
  \bibfield  {author} {\bibinfo {author} {\bibfnamefont {B.-F.}\ \bibnamefont
  {Li}}, \bibinfo {author} {\bibfnamefont {P.}~\bibnamefont {Singh}},\ and\
  \bibinfo {author} {\bibfnamefont {A.}~\bibnamefont {Wang}},\ }\bibfield
  {title} {\bibinfo {title} {{Phenomenological implications of modified loop
  cosmologies: an overview}},\ }\href
  {https://doi.org/10.3389/fspas.2021.701417} {\bibfield  {journal} {\bibinfo
  {journal} {Front. Astron. Space Sci.}\ }\textbf {\bibinfo {volume} {8}},\
  \bibinfo {pages} {701417} (\bibinfo {year} {2021})},\ \Eprint
  {https://arxiv.org/abs/2105.14067} {arXiv:2105.14067 [gr-qc]} \BibitemShut
  {NoStop}%
\bibitem [{\citenamefont {Zhang}\ \emph {et~al.}(2023)\citenamefont {Zhang},
  \citenamefont {Ma},\ and\ \citenamefont {Yang}}]{Zhang:2023okw}%
  \BibitemOpen
  \bibfield  {author} {\bibinfo {author} {\bibfnamefont {C.}~\bibnamefont
  {Zhang}}, \bibinfo {author} {\bibfnamefont {Y.}~\bibnamefont {Ma}},\ and\
  \bibinfo {author} {\bibfnamefont {J.}~\bibnamefont {Yang}},\ }\bibfield
  {title} {\bibinfo {title} {{Black hole image encoding quantum gravity
  information}},\ }\href {https://doi.org/10.1103/PhysRevD.108.104004}
  {\bibfield  {journal} {\bibinfo  {journal} {Phys. Rev. D}\ }\textbf {\bibinfo
  {volume} {108}},\ \bibinfo {pages} {104004} (\bibinfo {year} {2023})},\
  \Eprint {https://arxiv.org/abs/2302.02800} {arXiv:2302.02800 [gr-qc]}
  \BibitemShut {NoStop}%
\bibitem [{\citenamefont {Virbhadra}\ and\ \citenamefont
  {Ellis}(2000)}]{Virbhadra:1999nm}%
  \BibitemOpen
  \bibfield  {author} {\bibinfo {author} {\bibfnamefont {K.~S.}\ \bibnamefont
  {Virbhadra}}\ and\ \bibinfo {author} {\bibfnamefont {G.~F.~R.}\ \bibnamefont
  {Ellis}},\ }\bibfield  {title} {\bibinfo {title} {{Schwarzschild black hole
  lensing}},\ }\href {https://doi.org/10.1103/PhysRevD.62.084003} {\bibfield
  {journal} {\bibinfo  {journal} {Phys. Rev. D}\ }\textbf {\bibinfo {volume}
  {62}},\ \bibinfo {pages} {084003} (\bibinfo {year} {2000})},\ \Eprint
  {https://arxiv.org/abs/astro-ph/9904193} {arXiv:astro-ph/9904193}
  \BibitemShut {NoStop}%
\bibitem [{\citenamefont {Bozza}(2002)}]{Bozza:2002zj}%
  \BibitemOpen
  \bibfield  {author} {\bibinfo {author} {\bibfnamefont {V.}~\bibnamefont
  {Bozza}},\ }\bibfield  {title} {\bibinfo {title} {{Gravitational lensing in
  the strong field limit}},\ }\href
  {https://doi.org/10.1103/PhysRevD.66.103001} {\bibfield  {journal} {\bibinfo
  {journal} {Phys. Rev. D}\ }\textbf {\bibinfo {volume} {66}},\ \bibinfo
  {pages} {103001} (\bibinfo {year} {2002})},\ \Eprint
  {https://arxiv.org/abs/gr-qc/0208075} {arXiv:gr-qc/0208075} \BibitemShut
  {NoStop}%
\bibitem [{\citenamefont {Virbhadra}(2009)}]{Virbhadra:2008ws}%
  \BibitemOpen
  \bibfield  {author} {\bibinfo {author} {\bibfnamefont {K.~S.}\ \bibnamefont
  {Virbhadra}},\ }\bibfield  {title} {\bibinfo {title} {{Relativistic images of
  Schwarzschild black hole lensing}},\ }\href
  {https://doi.org/10.1103/PhysRevD.79.083004} {\bibfield  {journal} {\bibinfo
  {journal} {Phys. Rev. D}\ }\textbf {\bibinfo {volume} {79}},\ \bibinfo
  {pages} {083004} (\bibinfo {year} {2009})},\ \Eprint
  {https://arxiv.org/abs/0810.2109} {arXiv:0810.2109 [gr-qc]} \BibitemShut
  {NoStop}%
\bibitem [{\citenamefont {Tsukamoto}(2016)}]{Tsukamoto:2016qro}%
  \BibitemOpen
  \bibfield  {author} {\bibinfo {author} {\bibfnamefont {N.}~\bibnamefont
  {Tsukamoto}},\ }\bibfield  {title} {\bibinfo {title} {{Strong deflection
  limit analysis and gravitational lensing of an Ellis wormhole}},\ }\href
  {https://doi.org/10.1103/PhysRevD.94.124001} {\bibfield  {journal} {\bibinfo
  {journal} {Phys. Rev. D}\ }\textbf {\bibinfo {volume} {94}},\ \bibinfo
  {pages} {124001} (\bibinfo {year} {2016})},\ \Eprint
  {https://arxiv.org/abs/1607.07022} {arXiv:1607.07022 [gr-qc]} \BibitemShut
  {NoStop}%
\bibitem [{\citenamefont {Igata}(2025)}]{Igata:2025taz}%
  \BibitemOpen
  \bibfield  {author} {\bibinfo {author} {\bibfnamefont {T.}~\bibnamefont
  {Igata}},\ }\bibfield  {title} {\bibinfo {title} {{Deflection angle in the
  strong deflection limit: a perspective from local geometrical invariants and
  matter distributions}},\ }\href@noop {} {\  (\bibinfo {year} {2025})},\
  \Eprint {https://arxiv.org/abs/2503.02320} {arXiv:2503.02320 [gr-qc]}
  \BibitemShut {NoStop}%
\bibitem [{\citenamefont {Sahu}\ \emph {et~al.}(2015)\citenamefont {Sahu},
  \citenamefont {Lochan},\ and\ \citenamefont {Narasimha}}]{Sahu:2015dea}%
  \BibitemOpen
  \bibfield  {author} {\bibinfo {author} {\bibfnamefont {S.}~\bibnamefont
  {Sahu}}, \bibinfo {author} {\bibfnamefont {K.}~\bibnamefont {Lochan}},\ and\
  \bibinfo {author} {\bibfnamefont {D.}~\bibnamefont {Narasimha}},\ }\bibfield
  {title} {\bibinfo {title} {{Gravitational lensing by self-dual black holes in
  loop quantum gravity}},\ }\href {https://doi.org/10.1103/PhysRevD.91.063001}
  {\bibfield  {journal} {\bibinfo  {journal} {Phys. Rev. D}\ }\textbf {\bibinfo
  {volume} {91}},\ \bibinfo {pages} {063001} (\bibinfo {year} {2015})},\
  \Eprint {https://arxiv.org/abs/1502.05619} {arXiv:1502.05619 [gr-qc]}
  \BibitemShut {NoStop}%
\bibitem [{\citenamefont {Fu}\ and\ \citenamefont {Zhang}(2022)}]{Fu:2021fxn}%
  \BibitemOpen
  \bibfield  {author} {\bibinfo {author} {\bibfnamefont {Q.-M.}\ \bibnamefont
  {Fu}}\ and\ \bibinfo {author} {\bibfnamefont {X.}~\bibnamefont {Zhang}},\
  }\bibfield  {title} {\bibinfo {title} {{Gravitational lensing by a black hole
  in effective loop quantum gravity}},\ }\href
  {https://doi.org/10.1103/PhysRevD.105.064020} {\bibfield  {journal} {\bibinfo
   {journal} {Phys. Rev. D}\ }\textbf {\bibinfo {volume} {105}},\ \bibinfo
  {pages} {064020} (\bibinfo {year} {2022})},\ \Eprint
  {https://arxiv.org/abs/2111.07223} {arXiv:2111.07223 [gr-qc]} \BibitemShut
  {NoStop}%
\bibitem [{\citenamefont {Kumar}\ \emph {et~al.}(2023)\citenamefont {Kumar},
  \citenamefont {Islam},\ and\ \citenamefont {Ghosh}}]{Kumar:2023jgh}%
  \BibitemOpen
  \bibfield  {author} {\bibinfo {author} {\bibfnamefont {J.}~\bibnamefont
  {Kumar}}, \bibinfo {author} {\bibfnamefont {S.~U.}\ \bibnamefont {Islam}},\
  and\ \bibinfo {author} {\bibfnamefont {S.~G.}\ \bibnamefont {Ghosh}},\
  }\bibfield  {title} {\bibinfo {title} {{Strong gravitational lensing by loop
  quantum gravity motivated rotating black holes and EHT observations}},\
  }\href {https://doi.org/10.1140/epjc/s10052-023-12205-3} {\bibfield
  {journal} {\bibinfo  {journal} {Eur. Phys. J. C}\ }\textbf {\bibinfo {volume}
  {83}},\ \bibinfo {pages} {1014} (\bibinfo {year} {2023})},\ \Eprint
  {https://arxiv.org/abs/2305.04336} {arXiv:2305.04336 [gr-qc]} \BibitemShut
  {NoStop}%
\bibitem [{\citenamefont {Junior}\ \emph {et~al.}(2024)\citenamefont {Junior},
  \citenamefont {Lobo}, \citenamefont {Rodrigues},\ and\ \citenamefont
  {Vieira}}]{Junior:2023xgl}%
  \BibitemOpen
  \bibfield  {author} {\bibinfo {author} {\bibfnamefont {E.~L.~B.}\
  \bibnamefont {Junior}}, \bibinfo {author} {\bibfnamefont {F.~S.~N.}\
  \bibnamefont {Lobo}}, \bibinfo {author} {\bibfnamefont {M.~E.}\ \bibnamefont
  {Rodrigues}},\ and\ \bibinfo {author} {\bibfnamefont {H.~A.}\ \bibnamefont
  {Vieira}},\ }\bibfield  {title} {\bibinfo {title} {{Gravitational lens effect
  of a holonomy corrected Schwarzschild black hole}},\ }\href
  {https://doi.org/10.1103/PhysRevD.109.024004} {\bibfield  {journal} {\bibinfo
   {journal} {Phys. Rev. D}\ }\textbf {\bibinfo {volume} {109}},\ \bibinfo
  {pages} {024004} (\bibinfo {year} {2024})},\ \Eprint
  {https://arxiv.org/abs/2309.02658} {arXiv:2309.02658 [gr-qc]} \BibitemShut
  {NoStop}%
\bibitem [{\citenamefont {Li}\ and\ \citenamefont {Zhang}(2024)}]{Li:2024afr}%
  \BibitemOpen
  \bibfield  {author} {\bibinfo {author} {\bibfnamefont {H.}~\bibnamefont
  {Li}}\ and\ \bibinfo {author} {\bibfnamefont {X.}~\bibnamefont {Zhang}},\
  }\bibfield  {title} {\bibinfo {title} {{Gravitational Lensing Effects from
  Models of Loop Quantum Gravity with Rigorous Quantum Parameters}},\ }\href
  {https://doi.org/10.3390/universe10110421} {\bibfield  {journal} {\bibinfo
  {journal} {Universe}\ }\textbf {\bibinfo {volume} {10}},\ \bibinfo {pages}
  {421} (\bibinfo {year} {2024})}\BibitemShut {NoStop}%
\bibitem [{\citenamefont {Ashtekar}\ \emph
  {et~al.}(2018{\natexlab{b}})\citenamefont {Ashtekar}, \citenamefont
  {Olmedo},\ and\ \citenamefont {Singh}}]{Ashtekar:2018cay}%
  \BibitemOpen
  \bibfield  {author} {\bibinfo {author} {\bibfnamefont {A.}~\bibnamefont
  {Ashtekar}}, \bibinfo {author} {\bibfnamefont {J.}~\bibnamefont {Olmedo}},\
  and\ \bibinfo {author} {\bibfnamefont {P.}~\bibnamefont {Singh}},\ }\bibfield
   {title} {\bibinfo {title} {{Quantum extension of the Kruskal spacetime}},\
  }\href {https://doi.org/10.1103/PhysRevD.98.126003} {\bibfield  {journal}
  {\bibinfo  {journal} {Phys. Rev. D}\ }\textbf {\bibinfo {volume} {98}},\
  \bibinfo {pages} {126003} (\bibinfo {year} {2018}{\natexlab{b}})},\ \Eprint
  {https://arxiv.org/abs/1806.02406} {arXiv:1806.02406 [gr-qc]} \BibitemShut
  {NoStop}%
\bibitem [{\citenamefont {del Corral}\ and\ \citenamefont
  {Olmedo}(2022)}]{del-Corral:2022kbk}%
  \BibitemOpen
  \bibfield  {author} {\bibinfo {author} {\bibfnamefont {D.}~\bibnamefont {del
  Corral}}\ and\ \bibinfo {author} {\bibfnamefont {J.}~\bibnamefont {Olmedo}},\
  }\bibfield  {title} {\bibinfo {title} {{Breaking of isospectrality of
  quasinormal modes in nonrotating loop quantum gravity black holes}},\ }\href
  {https://doi.org/10.1103/PhysRevD.105.064053} {\bibfield  {journal} {\bibinfo
   {journal} {Phys. Rev. D}\ }\textbf {\bibinfo {volume} {105}},\ \bibinfo
  {pages} {064053} (\bibinfo {year} {2022})},\ \Eprint
  {https://arxiv.org/abs/2201.09584} {arXiv:2201.09584 [gr-qc]} \BibitemShut
  {NoStop}%
\bibitem [{\citenamefont {Motaharfar}\ and\ \citenamefont
  {Singh}(2025)}]{Motaharfar:2025typ}%
  \BibitemOpen
  \bibfield  {author} {\bibinfo {author} {\bibfnamefont {M.}~\bibnamefont
  {Motaharfar}}\ and\ \bibinfo {author} {\bibfnamefont {P.}~\bibnamefont
  {Singh}},\ }\bibfield  {title} {\bibinfo {title} {{Loop Quantum Gravitational
  Signatures via Love Numbers}},\ }\href@noop {} {\  (\bibinfo {year}
  {2025})},\ \Eprint {https://arxiv.org/abs/2501.09151} {arXiv:2501.09151
  [gr-qc]} \BibitemShut {NoStop}%
\bibitem [{\citenamefont {Ashtekar}\ and\ \citenamefont
  {Olmedo}(2020)}]{Ashtekar:2020ckv}%
  \BibitemOpen
  \bibfield  {author} {\bibinfo {author} {\bibfnamefont {A.}~\bibnamefont
  {Ashtekar}}\ and\ \bibinfo {author} {\bibfnamefont {J.}~\bibnamefont
  {Olmedo}},\ }\bibfield  {title} {\bibinfo {title} {{Properties of a recent
  quantum extension of the Kruskal geometry}},\ }\href
  {https://doi.org/10.1142/S0218271820500765} {\bibfield  {journal} {\bibinfo
  {journal} {Int. J. Mod. Phys. D}\ }\textbf {\bibinfo {volume} {29}},\
  \bibinfo {pages} {2050076} (\bibinfo {year} {2020})},\ \Eprint
  {https://arxiv.org/abs/2005.02309} {arXiv:2005.02309 [gr-qc]} \BibitemShut
  {NoStop}%
\bibitem [{\citenamefont {Cong}\ and\ \citenamefont
  {Zhang}(2020)}]{Zhang2020a}%
  \BibitemOpen
  \bibfield  {author} {\bibinfo {author} {\bibfnamefont {Z.}~\bibnamefont
  {Cong}}\ and\ \bibinfo {author} {\bibfnamefont {X.}~\bibnamefont {Zhang}},\
  }\bibfield  {title} {\bibinfo {title} {{Quantum geometry and effective
  dynamics of Janis-Newman-Winicour singularities}},\ }\href
  {https://doi.org/10.1103/PhysRevD.101.086002} {\bibfield  {journal} {\bibinfo
   {journal} {Phys. Rev. D}\ }\textbf {\bibinfo {volume} {101}},\ \bibinfo
  {pages} {086002} (\bibinfo {year} {2020})},\ \Eprint
  {https://arxiv.org/abs/1912.07278} {arXiv:1912.07278 [gr-qc]} \BibitemShut
  {NoStop}%
\bibitem [{\citenamefont {Duan}\ \emph {et~al.}(2023)\citenamefont {Duan},
  \citenamefont {Lin},\ and\ \citenamefont {Jia}}]{Duan:2023gvm}%
  \BibitemOpen
  \bibfield  {author} {\bibinfo {author} {\bibfnamefont {Y.}~\bibnamefont
  {Duan}}, \bibinfo {author} {\bibfnamefont {S.}~\bibnamefont {Lin}},\ and\
  \bibinfo {author} {\bibfnamefont {J.}~\bibnamefont {Jia}},\ }\bibfield
  {title} {\bibinfo {title} {{Deflection and gravitational lensing with finite
  distance effect in the strong deflection limit in stationary and axisymmetric
  spacetimes}},\ }\href {https://doi.org/10.1088/1475-7516/2023/07/036}
  {\bibfield  {journal} {\bibinfo  {journal} {JCAP}\ }\textbf {\bibinfo
  {volume} {07}},\ \bibinfo {pages} {036}},\ \Eprint
  {https://arxiv.org/abs/2304.07496} {arXiv:2304.07496 [gr-qc]} \BibitemShut
  {NoStop}%
\bibitem [{\citenamefont {Cao}\ \emph {et~al.}(2024)\citenamefont {Cao} \emph
  {et~al.}}]{LHAASO:2023uhj}%
  \BibitemOpen
  \bibfield  {author} {\bibinfo {author} {\bibfnamefont {Z.}~\bibnamefont
  {Cao}} \emph {et~al.} (\bibinfo {collaboration} {LHAASO}),\ }\bibfield
  {title} {\bibinfo {title} {{An ultrahigh-energy \ensuremath{\gamma}-ray
  bubble powered by a super PeVatron}},\ }\href
  {https://doi.org/10.1016/j.scib.2023.12.040} {\bibfield  {journal} {\bibinfo
  {journal} {Sci. Bull.}\ }\textbf {\bibinfo {volume} {69}},\ \bibinfo {pages}
  {449} (\bibinfo {year} {2024})},\ \Eprint {https://arxiv.org/abs/2310.10100}
  {arXiv:2310.10100 [astro-ph.HE]} \BibitemShut {NoStop}%
\bibitem [{\citenamefont {Zijian~Shi}\ and\ \citenamefont
  {Zhang}(2024)}]{Zhang2024c}%
  \BibitemOpen
  \bibfield  {author} {\bibinfo {author} {\bibfnamefont {Y.~M.}\ \bibnamefont
  {Zijian~Shi}}\ and\ \bibinfo {author} {\bibfnamefont {X.}~\bibnamefont
  {Zhang}},\ }\bibfield  {title} {\bibinfo {title} {{Higher-dimensional quantum
  Oppenheimer-Snyder model}},\ }\href
  {https://doi.org/10.1103/PhysRevD.110.104074} {\bibfield  {journal} {\bibinfo
   {journal} {Phys. Rev. D}\ }\textbf {\bibinfo {volume} {110}},\ \bibinfo
  {pages} {104074} (\bibinfo {year} {2024})},\ \Eprint
  {https://arxiv.org/abs/2408.15821} {arXiv:2408.15821 [gr-qc]} \BibitemShut
  {NoStop}%
\bibitem [{\citenamefont {Zhao}\ \emph {et~al.}(2024)\citenamefont {Zhao},
  \citenamefont {Tang},\ and\ \citenamefont {Xu}}]{Zhao:2024elr}%
  \BibitemOpen
  \bibfield  {author} {\bibinfo {author} {\bibfnamefont {L.}~\bibnamefont
  {Zhao}}, \bibinfo {author} {\bibfnamefont {M.}~\bibnamefont {Tang}},\ and\
  \bibinfo {author} {\bibfnamefont {Z.}~\bibnamefont {Xu}},\ }\bibfield
  {title} {\bibinfo {title} {{The Lensing Effect of Quantum-Corrected Black
  Hole and Parameter Constraints from EHT Observations}},\ }\href@noop {} {\
  (\bibinfo {year} {2024})},\ \Eprint {https://arxiv.org/abs/2403.18606}
  {arXiv:2403.18606 [gr-qc]} \BibitemShut {NoStop}%
\bibitem [{\citenamefont {Donoghue}\ \emph {et~al.}(2002)\citenamefont
  {Donoghue}, \citenamefont {Holstein}, \citenamefont {Garbrecht},\ and\
  \citenamefont {Konstandin}}]{Donoghue:2001qc}%
  \BibitemOpen
  \bibfield  {author} {\bibinfo {author} {\bibfnamefont {J.~F.}\ \bibnamefont
  {Donoghue}}, \bibinfo {author} {\bibfnamefont {B.~R.}\ \bibnamefont
  {Holstein}}, \bibinfo {author} {\bibfnamefont {B.}~\bibnamefont
  {Garbrecht}},\ and\ \bibinfo {author} {\bibfnamefont {T.}~\bibnamefont
  {Konstandin}},\ }\bibfield  {title} {\bibinfo {title} {{Quantum corrections
  to the Reissner-Nordstr\"om and Kerr-Newman metrics}},\ }\href
  {https://doi.org/10.1016/S0370-2693(02)01246-7} {\bibfield  {journal}
  {\bibinfo  {journal} {Phys. Lett. B}\ }\textbf {\bibinfo {volume} {529}},\
  \bibinfo {pages} {132} (\bibinfo {year} {2002})},\ \bibinfo {note} {[Erratum:
  Phys.Lett.B 612, 311--312 (2005)]},\ \Eprint
  {https://arxiv.org/abs/hep-th/0112237} {arXiv:hep-th/0112237} \BibitemShut
  {NoStop}%
\bibitem [{\citenamefont {Kirilin}(2007)}]{Kirilin:2006en}%
  \BibitemOpen
  \bibfield  {author} {\bibinfo {author} {\bibfnamefont {G.~G.}\ \bibnamefont
  {Kirilin}},\ }\bibfield  {title} {\bibinfo {title} {{Quantum corrections to
  the Schwarzschild metric and reparametrization transformations}},\ }\href
  {https://doi.org/10.1103/PhysRevD.75.108501} {\bibfield  {journal} {\bibinfo
  {journal} {Phys. Rev. D}\ }\textbf {\bibinfo {volume} {75}},\ \bibinfo
  {pages} {108501} (\bibinfo {year} {2007})},\ \Eprint
  {https://arxiv.org/abs/gr-qc/0601020} {arXiv:gr-qc/0601020} \BibitemShut
  {NoStop}%
\bibitem [{\citenamefont {Battista}(2024)}]{Battista:2023iyu}%
  \BibitemOpen
  \bibfield  {author} {\bibinfo {author} {\bibfnamefont {E.}~\bibnamefont
  {Battista}},\ }\bibfield  {title} {\bibinfo {title} {{Quantum Schwarzschild
  geometry in effective field theory models of gravity}},\ }\href
  {https://doi.org/10.1103/PhysRevD.109.026004} {\bibfield  {journal} {\bibinfo
   {journal} {Phys. Rev. D}\ }\textbf {\bibinfo {volume} {109}},\ \bibinfo
  {pages} {026004} (\bibinfo {year} {2024})},\ \Eprint
  {https://arxiv.org/abs/2312.00450} {arXiv:2312.00450 [gr-qc]} \BibitemShut
  {NoStop}%
\bibitem [{\citenamefont {Wang}\ and\ \citenamefont
  {Battista}(2025)}]{Wang:2025fmz}%
  \BibitemOpen
  \bibfield  {author} {\bibinfo {author} {\bibfnamefont {Z.-L.}\ \bibnamefont
  {Wang}}\ and\ \bibinfo {author} {\bibfnamefont {E.}~\bibnamefont
  {Battista}},\ }\bibfield  {title} {\bibinfo {title} {{Dynamical features and
  shadows of quantum Schwarzschild black hole in effective field theories of
  gravity}},\ }\href {https://doi.org/10.1140/epjc/s10052-025-13833-7}
  {\bibfield  {journal} {\bibinfo  {journal} {Eur. Phys. J. C}\ }\textbf
  {\bibinfo {volume} {85}},\ \bibinfo {pages} {304} (\bibinfo {year} {2025})},\
  \Eprint {https://arxiv.org/abs/2501.14516} {arXiv:2501.14516 [gr-qc]}
  \BibitemShut {NoStop}%
\bibitem [{\citenamefont {Maldacena}(1996)}]{Maldacena:1996ky}%
  \BibitemOpen
  \bibfield  {author} {\bibinfo {author} {\bibfnamefont {J.~M.}\ \bibnamefont
  {Maldacena}},\ }\emph {\bibinfo {title} {{Black holes in string theory}}},\
  \href@noop {} {Ph.D. thesis},\ \bibinfo  {school} {Princeton U.} (\bibinfo
  {year} {1996}),\ \Eprint {https://arxiv.org/abs/hep-th/9607235}
  {arXiv:hep-th/9607235} \BibitemShut {NoStop}%
\bibitem [{\citenamefont {Bena}\ \emph {et~al.}(2022)\citenamefont {Bena},
  \citenamefont {Martinec}, \citenamefont {Mathur},\ and\ \citenamefont
  {Warner}}]{Bena:2022rna}%
  \BibitemOpen
  \bibfield  {author} {\bibinfo {author} {\bibfnamefont {I.}~\bibnamefont
  {Bena}}, \bibinfo {author} {\bibfnamefont {E.~J.}\ \bibnamefont {Martinec}},
  \bibinfo {author} {\bibfnamefont {S.~D.}\ \bibnamefont {Mathur}},\ and\
  \bibinfo {author} {\bibfnamefont {N.~P.}\ \bibnamefont {Warner}},\ }\bibfield
   {title} {\bibinfo {title} {{Fuzzballs and Microstate Geometries: Black-Hole
  Structure in String Theory}},\ }\href@noop {} {\  (\bibinfo {year} {2022})},\
  \Eprint {https://arxiv.org/abs/2204.13113} {arXiv:2204.13113 [hep-th]}
  \BibitemShut {NoStop}%
\end{thebibliography}
\end{document}